\documentclass[12pt]{article}
\usepackage{epsfig}

\def\square{\kern1pt\vbox{\hrule height 1.2pt\hbox{\vrule width 1.2pt\hskip 3pt
   \vbox{\vskip 6pt}\hskip 3pt\vrule width 0.6pt}\hrule height 0.6pt}\kern1pt}
\begin{document}

\begin{titlepage}
\begin{flushright}
gr-qc/yymmnnn \\ CRETE-06-11 \\ UFIFT-QG-06-08
\end{flushright}

\vspace{0.5cm}

\begin{center}
\bf{A MAXIMALLY SYMMETRIC VECTOR PROPAGATOR}
\end{center}

\vspace{0.3cm}

\begin{center}
N. C. Tsamis$^{\dagger}$
\end{center}
\begin{center}
\it{Department of Physics, University of Crete \\
GR-710 03 Heraklion, HELLAS.}
\end{center}

\vspace{0.2cm}

\begin{center}
R. P. Woodard$^{\ddagger}$
\end{center}
\begin{center}
\it{Department of Physics, University of Florida \\
Gainesville, FL 32611, UNITED STATES.}
\end{center}

\vspace{0.3cm}

\begin{center}
ABSTRACT
\end{center}
\hspace{0.3cm} We derive the propagator for a massive vector field 
on a de Sitter background of arbitrary dimension. This propagator 
is de Sitter invariant and possesses the proper flat spacetime and 
massless limits. Moreover, the retarded Green's function inferred 
from it produces the correct classical response to a test source. 
Our result is expressed in a tensor basis which is convenient for 
performing quantum field theory computations using dimensional
regularization.

\vspace{0.3cm}

\begin{flushleft}
\end{flushleft}

\vspace{0.1cm}

\begin{flushleft}
$^{\dagger}$ e-mail: tsamis@physics.uoc.gr \\
$^{\ddagger}$ e-mail: woodard@phys.ufl.edu
\end{flushleft}

\end{titlepage}

\section{Introduction}

Vector fields described by the Proca Lagrangian \cite{AP}:
\begin{equation}
{\cal L} \; = \;
- \, \frac14 \, F_{\mu\nu} \, F_{\rho\sigma} \,
g^{\mu\rho} \, g^{\nu\sigma} \sqrt{-g} \; - \;
\frac12 \, m^2 \, A_{\mu} \, A_{\nu} \, g^{\mu\nu} \sqrt{-g}
\;\; , \label{L}
\end{equation}
have great physical relevance because they arise from spontaneous
symmetry breaking in gauge theories \cite{CW}. Although they have
been much studied in flat space, little attention has been given to
other geometries \cite{IMO}. In this paper we shall construct the Feynman
propagator of the Proca field described by (\ref{L}) on a $D$-dimensional
de Sitter background. Our motivation was to facilitate the dimensional
regularization computations in de Sitter background of the effective 
potential \cite{nctrpw1} and of loop scalar bilinear expectation values
\cite{PTW} in the theory of scalar quantum electrodynamics; taking the 
mass to zero gives the photon propagator in Lorentz gauge and that gauge 
turns out to effect great simplifications in perturbative calculations.
 
One might think the massive vector propagator on de Sitter background
was found some time ago in the classic work by Allen and Jacobson
\cite{allenjacobson}. However, their result diverges in the massless
limit, and fails to agree with the known flat space limit. The problem 
turns out to be a term missing from the right hand side of their propagator 
equation which only affects this one case of the many they considered. When 
the missing term is included the resulting solution is almost the same as 
what Allen and Jacobson got, but it remains finite in the massless limit 
and agrees with the known result in the flat space limit.

Section 2 derives the correct propagator equation. In Section 3 we introduce
a way of parameterizing the propagator that greatly facilitates perturbative
computations. The propagator equation is solved in Section 4. Section 5 gives 
four useful limits: the flat space limit, the coincidence limit, the massless 
limit, and the limit of $D = 4$ spacetime dimensions. Section 6 checks that 
the massless, retarded Green's function correctly reproduces the response to 
a point charge on the open coordinate sub-manifold. 

\section{The Propagator Equation}

The Proca equations associated with (\ref{L}) are:
\begin{equation}
\partial_{\nu} \Bigl( \sqrt{-g} \, g^{\nu\rho} \, g^{\mu\sigma} \,
F_{\rho\sigma}\Bigr) - m^2 \sqrt{-g} g^{\mu\nu} A_{\nu} \; = \; 0
\;\; . \label{eom}
\end{equation}
For non-zero mass, the Lorentz condition:
\begin{equation}
m^2 \neq 0 \qquad \Longrightarrow \qquad 
\partial_{\mu} \Bigl( \sqrt{-g}
g^{\mu\nu} \, A_{\nu}\Bigr) \; = \; 0 
\;\; . \label{supplement}
\end{equation}
is a direct consequence of the antisymmetry of the field strength
tensor:
\begin{equation}
F_{\mu\nu} \; = \;
\partial_{\mu} A_{\nu} - \partial_{\nu} A_{\mu}
\;\; . \label{Fmn}
\end{equation}
We shall continue to impose this condition even for $m^2 = 0$; in 
the massless case (\ref{L}) is gauge invariant and (\ref{supplement}) 
is a gauge condition. Using (\ref{supplement}) and employing covariant 
notation allows us to express the Proca equations in the form:
\begin{equation}
\mathcal{D}_{\mu}^{~\nu} A_{\nu} \; = \; 0 
\qquad , \qquad
{\cal D}_{\mu}^{~\nu} \; \equiv \; \sqrt{-g} \, \Bigl( \, 
\square_{\mu}^{~\nu} - R_{\mu}^{~\nu} - m^2 \delta_{\mu}^{~\nu} \Bigr)
\;\; . \label{quadratic}
\end{equation}
where $\mathcal{D}_{\mu}^{~\nu}$ represents the kinetic operator,
$\square_{\mu}^{~\nu}$ stands for the standard vector covariant 
d`Alembertian operator:
\begin{equation}
\square_{\mu}^{~\nu} A_{\nu} \; \equiv \; 
A_{\mu~\rho}^{~ ;\rho}
\;\; , \label{box}
\end{equation}
and $R_{\mu\nu}$ is the Ricci tensor.

The Feynman propagator should be the vacuum expectation value of the
time-ordered product of two vector fields:
\begin{equation}
i \Bigl[ {}_{\rho} \Delta_{\sigma} \Bigr](x;x') \; = \; \Bigl\langle \Omega
\Bigl\vert T\Bigl( A_{\rho}(x) A_{\sigma}(x')\Bigr) \Bigr\vert \Omega 
\Bigr\rangle \;\; . \label{fund}
\end{equation}
Assuming that the vacuum state is de Sitter invariant, the propagator must
be a bi-vector under de Sitter transformations, transforming according to 
$x^{\mu}$ on the index $\rho$ and according to $x^{\prime \mu}$ on the 
index $\sigma$. It would be simple enough to construct the propagator by 
working out the free field expansion for the vector potential and then taking 
the expectation value indicated in (\ref{fund}). We will instead follow
Allen and Jacobson \cite{allenjacobson} in searching for a de Sitter 
invariant solution to the differential equations the propagator must obey. 
If this solution is unique, it presumably agrees with the result of the 
canonical operator construction.

The propagator (\ref{fund}) must obviously obey the Lorentz condition
(\ref{supplement}) on both of its arguments:
\begin{eqnarray}
0 & = &
\partial_{\mu} \Bigl\{ \,
\sqrt{-g(x)} \, g^{\mu\nu}(x) \,\,
i \Bigl[ {}_{\nu} \Delta_{\rho} \Bigr](x;x') \, \Bigr\} 
\;\; , \nonumber \\
& = & 
\partial_{\rho}^{\hspace{0.05cm} \prime} \Bigl\{ \,
\sqrt{-g(x')} \, g^{\rho\sigma}(x') \,\,
i \Bigl[ {}_{\mu} \Delta_{\sigma} \Bigr](x;x') \, \Bigr\} 
\;\; . \label{gauge}
\end{eqnarray}
It must also invert, up to the factor of $-i$, the kinetic 
operator $\mathcal{D}_{\mu}^{~\nu}$ on the appropriate subspace 
of {\it transverse} functions. This automatically implies that:
\begin{equation}
\mathcal{D}_{\mu}^{~\rho} \, 
i \Bigl[ {}_{\rho} \Delta_{\nu} \Bigr](x;x') \; \neq \; 
i g_{\mu\nu} \delta^D(x - x')  
\;\; ,
\end{equation}
because the quantity on the left hand side is transverse on 
$x^{\prime\mu}$ whereas the quantity on the right hand side is 
not. The correct right hand side must be proportional to the 
transverse projection operator:
\begin{equation}
{\cal D}_{\mu}^{~\rho} \;
i \Bigl[ {}_{\rho} \Delta_{\nu} \Bigr](x;x') \; = \;
i g_{\mu\nu} \, \delta^D (x-x') \, + \,
\sqrt{-g(x)} \, \partial_{\mu} \, \partial^{\hspace{0.05cm} \prime}_{\nu} \;
i \Delta(x;x') \;\; , \label{propeqn}
\end{equation}
where $i \Delta(x;x')$ is the propagator of a massless, minimally
coupled scalar field in de Sitter spacetime. Transversality (\ref{gauge})
on $x^{\prime \mu}$ follows trivially using the equation for 
$i \Delta(x;x')$:
\begin{equation}
\partial_{\mu} \Bigl\{ \,
\sqrt{-g}(x) \, g^{\mu\nu}(x) \;
\partial_{\nu} \, i\Delta(x;x') \, \Bigr\} \; = \; 
i \delta^D(x - x') 
\;\; . \label{scalprop}
\end{equation}

It should be noted that the correct vector propagator equation 
(\ref{propeqn}) is nothing but the extension to curved spacetimes 
of the familiar flat spacetime relation:
\begin{equation}
\Bigl( \partial^2 - m^2 \Bigr) \;
i\Bigl[ {}_{\mu} \Delta_{\nu} \Bigr](x;x') \; = \;
i\left( \eta_{\mu\nu} - \frac{\partial_{\mu} \, 
\partial_{\nu}}{\partial^2} \right) \, \delta^D (x-x') 
\;\; , \label{propeqnflat}
\end{equation}
The extra term on the right hand side of this relation, and of 
(\ref{propeqn}), is clearly essential for transversality. However, 
it is just as clearly non-zero away from coincidence, that is, for 
$x^{\prime\mu} \neq x^{\mu}$. The technique of Allen and Jacobson 
was to solve the various propagator equations away from coincidence, 
with zero on the right hand side. They then adjusted the coefficients 
of the linearly independent, homogeneous solutions to enforce two 
conditions \cite{allenjacobson}:
\begin{itemize}
\item{Analyticity at the antipodal point, $x^{\prime \mu} = 
\overline{x}^{\mu}$; and}
\item{The proper singularity structure at coincidence to result in 
a delta function on the right hand side.}
\end{itemize}
This technique is completely valid when the right hand side contains 
only a delta function, as is the case when a gauge fixing term is 
added to the action. However, it fails for an exact gauge such as 
(\ref{supplement}), for which the right hand side of the propagator 
equation must project onto the subspace of functions respecting the 
gauge condition. In that case the right hand side of the propagator 
equation is generally non-zero, even away from coincidence, and the 
propagator must include an inhomogeneous solution to reproduce this 
non-zero term. That is the only thing Allen and Jacobson missed.

\section{The $y$-Basis}

Although much of our analysis will be independent of the coordinate 
system, one should think in terms of conformally flat coordinates
for which the invariant line element is:
\begin{equation}
ds^2 \, = \,
a^2(\eta) \, \Big( -d\eta^2 \, + \,
d{\vec x} \cdot d{\vec x} \, \Big)
\;\; , \label{ds2}
\end{equation}
where $a(\eta)$ and $H$ are the scale factor and Hubble parameter 
respectively:
\footnote{In (\ref{H}) $\Lambda$ is the cosmological constant.}
\begin{equation}
a(\eta) \, = \,
- \frac{1}{H \eta}
\quad , \quad
H^2 \, \equiv \, \frac{1}{D-1} \, \Lambda \, > \, 0
\;\; . \label{H}
\end{equation}

Although any scalar function of $x^{\mu}$ and $x^{\prime\mu}$ could 
be expressed in terms of the geodesic length $\ell(x \, ; x')$, it 
is often simpler to use the following length function:
\begin{equation}
y(x \, ; x') \; \equiv \;
H^2 a(\eta) \, a(\eta') \, \Delta x^2(x \, ; x')
\;\; , \label{y}
\end{equation}
where we have defined:
\footnote{The infinitesimal imaginary part of (\ref{deltax}) 
enforces Feynman boundary conditions.}
\begin{equation}
\Delta x \; \equiv \;
\sqrt{\Vert {\vec x} - {\vec x}' \Vert^2 \, - \,
(\vert \eta - \eta' \vert - i\varepsilon)^2}
\;\; . \label{deltax}
\end{equation}
The relation between $y(x \, ; x')$ and $\ell(x \, ; x')$ is:
\begin{equation}
y(x \, ; x') = 4 \sin^2\Biggl( \frac12 H \ell(x \, ; x')\Biggr) 
\;\; . \label{ytoell}
\end{equation}
One can see from (\ref{ytoell}) that the length function between 
a point and its antipodal is:
\begin{equation}
y(x;\overline{x}) \; = \; 4
\;\; .
\end{equation}

It has long been realized that any de Sitter invariant bi-vector can be 
expressed in terms of the geodesic length $\ell(x;x')$, the gradients of
$\ell(x;x')$ with respect to $x^{\mu}$ and $x^{\prime \mu}$, and the parallel 
transport matrix $[\mbox{}_{\mu} g_{\nu}](x;x')$ \cite{allenjacobson,peters}. 
However, this is not convenient for theories such as scalar quantum 
electrodynamics and gravity which possess derivative interactions. The more 
convenient tensor basis consists simply of de Sitter invariant derivatives 
of $y(x;x')$. Using this basis, derivatives of functions of $y(x;x')$ 
immediately produce basis tensors, and one can dispense with the tedious 
step of expressing these derivatives in terms of gradients of $\ell(x;x')$ 
-- using relation (\ref{ytoell}) -- and the parallel transport matrix --
using the identity \cite{KW}:
\begin{equation}
\Bigl[ {}_{\mu} g_{\nu} \Bigr](x;x') \; = \; -\frac1{2 H^2} 
\; \frac{\partial^2 y}{\partial x^{\mu} \, 
\partial x^{\prime \hspace{0.05cm} \nu}} - \, \frac1{2 H^2 (4 - y)} \,
\frac{\partial y}{\partial x^{\mu}} \, 
\frac{\partial y}{\partial x^{\prime \hspace{0.05cm} \nu}} \;\; .
\end{equation}

It is simple to work out the contraction of any two basis tensors in
conformal coordinates. Because these contractions are de Sitter invariant,
the relations one obtains in this fashion must then apply in any coordinate
system. The contractions we require are \cite{KW}:
\begin{eqnarray}
g^{\mu\nu}(x) \; \frac{\partial y}{\partial x^{\mu}} 
\frac{\partial y}{\partial x^{\nu}} & = & H^2 \Bigl(4 y - y^2\Bigr) 
\;\; , \label{ID1} \\
g^{\mu\nu}(x) \; \frac{\partial y}{\partial x^{\mu}} 
\frac{\partial^2 y}{\partial x^{\nu} \partial x^{\prime \rho}} 
& = & H^2 (2-y) \, \frac{\partial y}{\partial x^{\prime \rho}} 
\;\; , \label{ID2} \\
g^{\mu\nu}(x) \; \frac{\partial^2 y}{\partial x^{\mu} \partial x^{\prime \rho}}
\frac{\partial^2 y}{\partial x^{\nu} \partial x^{\prime \sigma}} 
& = & 4 H^4 \,
g_{\rho\sigma}(x') \, - \, H^2 \frac{\partial y}{\partial x^{\prime \rho}}
\frac{\partial y}{\partial x^{\prime \sigma}} 
\;\; . \label{ID3}
\end{eqnarray}
We will also need covariant derivatives of the basis tensors \cite{KW}:
\begin{eqnarray}
\frac{D}{D x^{\mu}} \, \frac{\partial y}{\partial x^{\nu}} & \equiv & 
\frac{\partial^2 y}{ \partial x^{\mu} \partial x^{\nu}} \, - \,
\Gamma^{\rho}_{~\mu\nu}(x) \, \frac{\partial y}{\partial x^{\rho}} 
\; = \; H^2 (2-y) \, g_{\mu\nu}(x) 
\;\; , \label{ID4} \\
\square_{\mu}^{~\nu} \, \frac{\partial^2 y}
{\partial x^{\nu} \partial x^{\prime \rho}} & = & 
-H^2 \frac{\partial^2 y}{\partial x^{\mu} \partial x^{\prime \rho}} 
\;\; . \label{ID5}
\end{eqnarray}
Each of these identities (\ref{ID1}-\ref{ID5}) applies as well 
when $x^{\mu}$ and $x^{\prime \mu}$ are interchanged.

One can express any de Sitter invariant vector propagator as follows 
in the $y$-basis \cite{KW}:
\begin{equation}
i \Bigl[ {}_{\mu} \Delta_{\nu} \Bigr](x;x') \; = \;
B(y) \; \frac{\partial^2 y}{\partial x^{\mu} \,
\partial x^{\prime \hspace{0.05cm} \nu}}
\, + \,
C(y) \; \frac{\partial y}{\partial x^{\mu}} \,
\frac{\partial y}{\partial x^{\prime \hspace{0.05cm} \nu}}
\;\; . \label{prop}
\end{equation}
Substituting this representation into the transversality relation
(\ref{gauge}) and making use of identities (\ref{ID1}-\ref{ID5}) gives:
\begin{equation}
-D \, B(y) \, + \, (2-y) \, B^{\prime}(y) \, + \,
(D+1)(2-y) \, C(y) \, + \, (4y-y^2) \, C^{\prime}(y)
\; = \; 0
\;\; . \label{transv}
\end{equation}
This equation implies that the functions $B(y)$ and $C(y)$ can be
expressed in terms of a single function $\gamma(y)$:
\begin{eqnarray}
B(y) & = & \frac{1}{4(D-1)H^2} \,
\Bigl[ -(4y-y^2) \frac{\partial}{\partial y} \, - \,
(D-1)(2-y) \Bigr] \, \gamma(y)
\;\; , \label{B} \\
C(y) & = & \frac{1}{4(D-1)H^2} \,
\Bigl[ (2-y) \frac{\partial}{\partial y} \, - \,
(D-1) \Bigr] \, \gamma(y)
\;\; . \label{C}
\end{eqnarray}
Relations (\ref{prop}-\ref{C}) are independent of the propagator 
equation (\ref{propeqn}). In fact these formulae are simply the 
translation into the $y$-basis of results previously obtained by 
Allen and Jacobson \cite{allenjacobson}.

We close this section by expressing the propagator equation (\ref{propeqn}) 
in the $y$-basis, beginning with the right hand side. It has long been 
recognized that equation (\ref{scalprop}) for the scalar propagator has no 
de Sitter invariant solution \cite{allenfolacci}. However, the mixed 
derivative $\partial_{\mu} \partial_{\nu}' i\Delta(x;x')$ which appears in 
the vector propagator equation (\ref{propeqn}) is nonetheless de Sitter 
invariant, up to a delta function term.
\footnote{For example, it is the coincidence limit of this mixed 
derivative that occurs in the de Sitter invariant expectation value 
of the scalar stress tensor \cite{allenfolacci}.} 
If we elect to maintain homogeneity and isotropy, the scalar propagator 
takes the form \cite{OW1}:
\begin{equation}
i\Delta(x;x') \; = \;
A(y) \, + \, k \ln \Bigl[ a(\eta) \, a(\eta') \Bigr]
\;\; . \label{DeltaA}
\end{equation}
Therefore, the extra term on the right hand side of (\ref{propeqn}) is:
\begin{eqnarray}
\lefteqn{\sqrt{-g(x)} \, \partial_{\mu} \, 
\partial^{\hspace{0.05cm} \prime}_{\nu} \; 
i\Delta (x;x') \; = \; 
\delta^0_{\mu} \, \delta^0_{\nu} \, i \delta^D(x - x') } 
\nonumber \\
& & \hspace{3cm} 
+ \sqrt{-g(x)} \left[ A^{\prime\prime}(y) \;
\frac{\partial y}{\partial x^{\mu}} \,
\frac{\partial y}{\partial x^{\prime \hspace{0.05cm} \nu}}
\; + \;
A^{\prime}(y) \;
\frac{\partial^2 y}{\partial x^{\mu} \,
\partial x^{\prime \hspace{0.05cm} \nu}}
\right]
\;\; . \qquad \label{Acontr}
\end{eqnarray}
The de Sitter breaking logarithms of (\ref{DeltaA}) drop out in 
the mixed derivative, and we need not worry about the delta function 
terms if we follow Allen and Jacobson in solving the equation away 
from coincidence.

Before going on to the vector propagator it is worth pointing out 
a few more things about the scalar propagator. First, its equation 
(\ref{scalprop}) implies the following relation for the function 
$A(y)$ for $y \neq 0$ \cite{OW1}:
\begin{equation}
(4y-y^2) \, A^{\prime\prime}(y) \, + \,
D(2-y) \, A^{\prime}(y) \; = \;
(D-1) \, k
\;\; , \label{A''eqn}
\end{equation}
Second, the de Sitter invariant function $A(y)$ is \cite{OW2}:
\begin{eqnarray}
A(y) & = & \frac{H^{D-2}}{(4\pi)^{\frac{D}{2}}} \;
\Bigg\{ \,
\frac{\Gamma(\frac{D}{2})}{(\frac{D}{2}-1)} \,
\left( \frac{4}{y} \right)^{\frac{D}{2}-1} + \,
\frac{\Gamma(\frac{D}{2}+1)}{(\frac{D}{2}-2)} \,
\left( \frac{4}{y} \right)^{\frac{D}{2}-2}
\nonumber \\
& \mbox{} & \hspace{4.9cm}
- \; \pi \cot (\frac{\pi}{2} D) \;
\frac{\Gamma(D-1)}{\Gamma(\frac{D}{2})}
\nonumber \\
& \mbox{} &
+ \, \sum_{n=0}^{\infty} \Bigg[ \,
\frac{\Gamma(\frac{D}{2}+1+n)}{\Gamma(n+2)} \,
\frac{ \left( \frac{y}{4} \right)^{n-\frac{D}{2}+2} }
{ \frac{D}{2}-2-n } \; + \,
\frac{\Gamma(D-1+n)}{\Gamma(\frac{D}{2}+n)} \,
\frac{ \left( \frac{y}{4} \right)^n }{n} \, \Bigg]
\, \Bigg\} 
\;\; , \qquad \label{A}
\end{eqnarray}
and the constant $k$ has the value \cite{OW1}:
\begin{equation}
k \, \equiv \, \frac{H^{D-2}}{(4\pi)^{\frac{D}{2}}} \,
\frac{\Gamma(D-1)}{\Gamma(\frac{D}{2})} 
\;\; . \label{k}
\end{equation}
Finally, the derivative of $A(y)$ can be expressed in terms of
a hypergeometric function \cite{GR}:
\begin{equation}
A^{\prime}(y) \; = \;
- \frac14 \; \frac{H^{D-2}}{(4\pi)^{\frac{D}{2}}} \;
\frac{\Gamma(D) \, \Gamma(1)}{\Gamma(\frac{D}{2}+1)} \;\;
{}_{2}F_{1} \left( D \; , \, 1 \; , \, \frac{D}{2} + 1 \; ; \,
1 - \frac{y}{4} \right)
\;\; . \label{A'}
\end{equation}

We are now ready to tackle the left hand side of the propagator equation
(\ref{propeqn}). By substituting (\ref{prop}) and (\ref{Acontr}), and again 
making use of (\ref{ID1}-\ref{ID5}), we obtain two differential equations 
involving $B(y)$ and $C(y)$ away from the singular point $y=0$. The term 
proportional to $(\partial^2 y / \partial x \, \partial x^{\prime})$ implies:
\begin{equation}
(4y-y^2) \, B^{\prime\prime} \, + \,
D(2-y) \, B^{\prime} \, - \,
\left( D + \frac{m^2}{H^2} \right) B \, + \,
2(2-y) \, C \; = \;
\frac{1}{H^2} \, A^{\prime}
\;\; , \label{B''eqn}
\end{equation}
while the term proportional to $(\partial y / \partial x) \,
(\partial y / \partial x^{\prime})$ leads to:
\begin{equation}
(4y-y^2) \, C^{\prime\prime} \, + \,
(D+4)(2-y) \, C^{\prime} \, - \,
\left( 2D + \frac{m^2}{H^2} \right) C \, - \,
2 \, B^{\prime} \; = \;
\frac{1}{H^2} \, A^{\prime\prime}
\;\; . \label{C''eqn}
\end{equation}
Now $B(y)$ and $C(y)$ are not independent variables; they are
related to one another by the transversality constraint (\ref{transv}).
Using (\ref{transv}) and (\ref{A''eqn}), it is straightforward to 
see that the following combination of equations (\ref{B''eqn}) and 
(\ref{C''eqn}) gives a tautology of the form $0 = 0$:
\begin{eqnarray}
(2 - y) \, \frac{\partial}{\partial y} (eqn\ref{B''eqn}) \, + \,
(4y - y^2) \, \frac{\partial}{\partial y} (eqn\ref{C''eqn}) \, - \,
D \, (eqn\ref{B''eqn}) 
\nonumber \\
& \mbox{} & \hspace{-4cm}
\, + \, (D+1) (2-y) \, (eqn\ref{C''eqn}) 
\;\; .
\end{eqnarray}

The independent combination of (\ref{B''eqn}) and (\ref{C''eqn}) can be
obtained by inverting relations (\ref{B}) and (\ref{C}) to express
$\gamma(y)$ as:
\begin{equation}
\gamma(y) \; = \;
-H^2 \Bigl[ (2-y) \, B(y) \, + \,
(4y-y^2) \, C(y) \Bigr]
\;\; . \label{gamma}
\end{equation}
Its derivatives equal:
\begin{eqnarray}
\gamma^{\prime} & = &
-H^2 \Bigl[ (2-y)B^{\prime} \, - \, B \, + \,
(4y-y^2)C^{\prime} \, + \, 2(2-y)C \Bigr]
\; , \label{gamma'} \\
\gamma^{\prime\prime} & = &
-H^2 \Bigl[ (2-y)B^{\prime\prime} \, - \,
2B^{\prime} \, + \, (4y-y^2)C^{\prime\prime} \, + \,
4(2-y)C^{\prime} \, - \, 2C \Bigr]
\; . \qquad \label{gamma''}
\end{eqnarray}
Now we make use of (\ref{B''eqn}), (\ref{C''eqn}) and (\ref{A''eqn}) 
to infer:
\begin{eqnarray}
(4y-y^2) \, \gamma^{\prime\prime}(y) \, + \,
(D+2)(2-y) \, \gamma^{\prime}(y) \, - \,
\left[ 2(D-1) \, + \, \frac{m^2}{H^2} \right] \gamma(y)
& \mbox{} & \nonumber \\
&\mbox{} & \hspace{-8.6cm} = \;
- (4y-y^2) \, A^{\prime\prime}(y) \, - \,
(2-y) \, A^{\prime}(y)
\;\; , \label{gammaeqn} \\
&\mbox{} & \hspace{-8.6cm}
= \; (D-1) \, \Bigl[ (2-y) \, A^{\prime}(y) \, - \, k \Bigr]
\;\; . \label{gammaeqn2}
\end{eqnarray}
This is the fundamental expression of the propagator equation 
(\ref{propeqn}) in terms of the $y$-basis.

\section{The Solution}

The general solution of equation (\ref{gammaeqn2}) takes the form:
\begin{equation}
\gamma(y) \; = \;
c_1 \gamma_1(y) \, + \, c_2 \gamma_2(y) \, + \, \gamma_p(y)
\;\; , \label{gensol}
\end{equation}
where $\gamma_1(y)$ and $\gamma_2(y)$ are the linearly independent
solutions of the homogeneous version of equation (\ref{gammaeqn2}) and 
$\gamma_p(y)$ a particular solution of the full, inhomogeneous equation. 
From (\ref{A''eqn}) it is straightforward to see that a particular 
solution is:
\begin{equation}
\gamma_p(y) \; = \; \frac{H^2}{m^2} \, 
(D-1) \Bigl[ - (2-y) A^{\prime}(y) \, - \, k \Bigr]
\;\; .
\end{equation}
Using relation (\ref{A'}) and a few simple facts about hypergeometric
functions \cite{GR}, we can write $\gamma_p(y)$ explicitly as:
\begin{equation}
\gamma_p(y) \; = \; \frac{H^2}{m^2} \, 
\frac{H^{D-2}}{(4\pi)^{\frac{D}{2}}} \,
\frac{\Gamma(D)}{2 \Gamma(\frac{D}2 + 1)} \;\;
{}_{2} F_{1} \left( D-1 \, , \, 2 \, , \, \frac{D}{2} + 1 \, ; \,
1 - \frac{y}{4} \right) 
\;\; . \label{partsol}
\end{equation}
The two homogeneous solutions can be similarly expressed as hypergeometric 
functions \cite{GR}:
\begin{eqnarray}
\gamma_1(y) & = &
{}_{2} F_{1}
\left( \frac{D+1}{2} + \nu \; , \, \frac{D+1}{2} - \nu \; , \,
\frac{D}{2} + 1 \; ; \, 1 - \frac{y}{4} \right)
\;\; , \label{homsol1} \\
\gamma_2(y) & = & {}_{2} F_{1} \left( \frac{D+1}{2} + \nu \; , \,
\frac{D+1}{2} - \nu \; , \, \frac{D}{2} + 1 \; ; \, \frac{y}{4}
\right)
\;\; , \label{homsol2}
\end{eqnarray}
where we have defined:
\begin{equation}
\nu \; \equiv \;
\sqrt{ \left( \frac{D-3}{2} \right)^2 \, - \,
\frac{m^2}{H^2} }
\;\; . \label{nu}
\end{equation}
We must now determine the two arbitrary constants $c_1$ and $c_2$
present in (\ref{gensol}) by enforcing the proper behaviour at
$y=0$ and at $y=4$.

The desired behavior at $y = 4$ -- that is, at the antipodal point
-- is for the solution to be analytic. This is most easily examined
by employing the appropriate expansion of the hypergeometric
function \cite{GR}:
\begin{eqnarray}
{}_{2} F_{1}
\left( \alpha \; , \, \beta \; , \, \delta \; ; \, z \right)
& = & \sum_{n=0}^{\infty}
\frac{\Gamma(\alpha+n) \, \Gamma(\beta+n) \, \Gamma(\delta)}
{\Gamma(\alpha) \, \Gamma(\beta) \, \Gamma(\delta+n)} \;
\frac{z^n}{n!}
\;\; , \label{F} \\
& = &
1 \, + \, \frac{\alpha\beta}{\delta} \, z \, + \,
\frac{\alpha (\alpha+1) \, \beta (\beta+1)}{\delta (\delta+1)} \,
\frac{z^2}{2!} \, + \, {\cal O}(z^3)
\;\; . \qquad \label{Fexp}
\end{eqnarray}
It is apparent that $\gamma_p(y)$ -- given by (\ref{partsol})
-- and $\gamma_1(y)$ -- given by (\ref{homsol1}) -- are analytic
at $y = 4$ while $\gamma_2(y)$ -- given by (\ref{homsol2}) -- is
not. We must, therefore, set:
\begin{equation}
c_2 \, = \, 0
\qquad \Longrightarrow \qquad
\gamma(y) \; = \; c_1 \gamma_1(y) \, + \, \gamma_p(y)
\;\; . \label{c2}
\end{equation}

The desired behavior at $y=0$ -- that is, at coincidence -- is for
the solution to be $\sim y^{1-\frac{D}{2}}$. This is achieved by
enforcing the following condition on (\ref{c2}):
\begin{equation}
\Bigl[ \gamma \Bigr]_{(\frac{4}{y})^{\frac{D}{2}}} \; = \;
c_1 \Bigl[ \gamma_1 \Bigr]_{(\frac{4}{y})^{\frac{D}{2}}} \, + \,
\Bigl[ \gamma_p \Bigr]_{(\frac{4}{y})^{\frac{D}{2}}} \; = \; 0 
\;\; , \label{c1eqn}
\end{equation}
where the subscript on the brackets in (\ref{c1eqn}) indicates
the term in the Laurent expansion of the bracketed quantity which is
proportional to the subscript. To determine this part we use one of
the transformation formulae of hypergeometric functions \cite{GR}:
\begin{eqnarray}
{}_{2} F_{1}
\left( \alpha \; , \, \beta \; , \, \delta \; ; \, z \right)
& = &
\frac{\Gamma(\delta) \, \Gamma(\delta-\alpha-\beta)}
{\Gamma(\delta-\alpha) \, \Gamma(\delta-\beta)} \;\;
{}_{2} F_{1}
\left( \alpha \; , \, \beta \; , \, \alpha+\beta-\delta+1
\; ; \, 1-z \right)
\nonumber \\
& \mbox{} & \hspace{-2cm}
+ \, \frac{\Gamma(\delta) \, \Gamma(\alpha+\beta-\delta)}
{\Gamma(\alpha) \, \Gamma(\beta)} \;
(1-z)^{\delta-\alpha-\beta}
\nonumber \\
& \mbox{} &
\times \;\; {}_{2} F_{1}
\left( \delta-\alpha \; , \, \delta-\beta \; , \, \delta-\alpha-\beta+1
\; ; \, 1-z \right)
\; \; . \label{Ftransf}
\end{eqnarray}
Applying (\ref{Ftransf}) to our particular values:
\begin{equation}
\alpha \; = \; \frac{D+1}{2} + \nu
\quad , \quad
\beta \; = \; \frac{D+1}{2} - \nu
\quad , \quad
\delta \; = \; \frac{D}{2} + 1
\;\; , \label{abd}
\end{equation}
gives for the relevant part of our hypergeometric function:
\begin{eqnarray}
& \mbox{} & \hspace{-2.3cm}
\Biggl[ {}_{2} F_{1}
\left( \frac{D+1}{2} + \nu \; , \, \frac{D+1}{2} - \nu \; , \,
\frac{D}{2} + 1 \; ; \, 1 - \frac{y}{4} \right)
\Biggr]_{(\frac{4}{y})^{\frac{D}{2}}}
\nonumber \\
& \mbox{} & \hspace{3.3cm}
= \; \frac{\Gamma(\frac{D}{2} + 1) \, \Gamma(\frac{D}{2})}
{\Gamma(\frac{D+1}{2} + \nu) \, \Gamma(\frac{D+1}{2} - \nu)} \;
\left( \frac{4}{y} \right)^{\frac{D}{2}}
\;\; . \label{F(4/y)}
\end{eqnarray}
Moreover, using (\ref{A'}) we also conclude:
\begin{equation}
\Bigl[ A' \Bigr]_{(\frac{4}{y})^{\frac{D}{2}}}
\; = \;
- \frac14 \; \frac{H^{D-2}}{(4\pi)^{\frac{D}{2}}} \;
\Gamma(\frac{D}{2}) \; \left( \frac{4}{y} \right)^{\frac{D}{2}}
\;\; . \label{A'(4/y)}
\end{equation}
Inserting (\ref{F(4/y)}-\ref{A'(4/y)}) into (\ref{c1eqn})
reveals that the latter can only be satisfied if:
\begin{equation}
c_1 \; = \;
- \frac{D-1}{2} \; \frac{H^2}{m^2} \;
\frac{H^{D-2}}{(4\pi)^{\frac{D}{2}}} \;
\frac{\Gamma(\frac{D+1}{2} + \nu) \, \Gamma(\frac{D+1}{2} - \nu)}
{\Gamma(\frac{D}{2} + 1)}
\;\; . \label{c1}
\end{equation}

Having determined both constants $c_1$ and $c_2$, we arrive at the
unique solution:
\begin{eqnarray}
\gamma(y) &\!\!\! = \!\!\!&
- \frac{D-1}{2} \, \frac{H^2}{m^2} \,
\frac{H^{D-2}}{(4\pi)^{\frac{D}{2}}} \,
\Bigg\{
\!- \frac{\Gamma(D-1)}{\Gamma(\frac{D}{2} + 1)} \;\;
{}_{2} F_{1} \left( D-1 \, , \, 2 \, , \, \frac{D}{2} + 1 \, ; \,
1 - \frac{y}{4} \right)
\nonumber \\
& \mbox{} & \hspace{1.7cm}
+ \, \frac{\Gamma(\frac{D+1}{2} + \nu) \,
\Gamma(\frac{D+1}{2} - \nu)}
{\Gamma(\frac{D}{2} + 1)}
\label{sol} \\
& \mbox{} & \hspace{2.5cm}
\times \; {}_{2} F_{1} \left(
\frac{D+1}{2} + \nu \, , \, \frac{D+1}{2} - \nu \, , \,
\frac{D}{2} + 1 \, ; \, 1 - \frac{y}{4} \right) \Bigg\}
\;\; , \nonumber
\end{eqnarray}
where the parameter $\nu$ is defined in (\ref{nu}). The second 
hypergeometric function in (\ref{sol}) represents the solution
obtained by Allen and Jacobson \cite{allenjacobson}; the first
hypergeometric function is our contribution. It is often useful
to have the Laurent expansion of (\ref{sol}):
\begin{eqnarray}
\gamma(y) & = &
- \frac{D-1}{2} \, \frac{H^2}{m^2} \,
\frac{H^{D-2}}{(4\pi)^{\frac{D}{2}}} \,
\Bigg\{ \!- \frac{m^2}{H^2} \; \Gamma(\frac{D}{2} - 1) \,
\left( \frac{4}{y} \right)^{\frac{D}{2} - 1}
\nonumber \\
& \mbox{} &
+ \, \sum_{n=0}^{\infty} \Bigg[
-(n+1) \, \frac{\Gamma(n+D-1)}{\Gamma(n+\frac{D}{2}+1)} \,
\left( \frac{y}{4} \right)^{n}
\nonumber \\
& \mbox{} & \hspace{2cm}
+ \, (n-\frac{D}{2}+3) \,
\frac{\Gamma(n+\frac{D}{2}+1)}{\Gamma(n+3)} \,
\left( \frac{y}{4} \right)^{n-\frac{D}{2}+2}
\Bigg]
\nonumber \\
& \mbox{} & \hspace{-1cm}
+ \, \frac{\Gamma(\frac{D}{2}-1) \, \Gamma(2-\frac{D}{2})}
{\Gamma(\frac12+\nu) \, \Gamma(\frac12-\nu)} \,
\sum_{n=0}^{\infty} \Bigg[ \,
\frac{\Gamma(n+\frac{D+1}{2}+\nu) \, \Gamma(n+\frac{D+1}{2}-\nu)}
{\Gamma(n+\frac{D}{2}+1) \, \Gamma(n+1)} \,
\left( \frac{y}{4} \right)^n
\nonumber \\
& \mbox{} & \hspace{1.5cm}
- \, \frac{\Gamma(n+\frac{5}{2}+\nu) \, \Gamma(n+\frac{5}{2}-\nu)}
{\Gamma(n+3) \, \Gamma(n-\frac{D}{2}+3)} \,
\left( \frac{y}{4} \right)^{n-\frac{D}{2}+2}
\Bigg] \Bigg\}
\;\; . \label{sol3}
\end{eqnarray}

\section{Various Limits}

In this section we derive interesting limits for the function $\gamma(y)$ 
whose alternate expressions, (\ref{sol}) and (\ref{sol3}), have just been 
given. Where appropriate we also derive limits for the full propagator, 
whose dependence upon $\gamma(y)$ is restated here for convenience:
\begin{eqnarray}
\lefteqn{i \Bigl[ {}_{\mu} \Delta_{\nu} \Bigr](x;x') \; = \;
B(y) \; \frac{\partial^2 y}{\partial x^{\mu} \,
\partial x^{\prime \hspace{0.05cm} \nu}}
\, + \,
C(y) \; \frac{\partial y}{\partial x^{\mu}} \,
\frac{\partial y}{\partial x^{\prime \hspace{0.05cm} \nu}}
\;\; , } \label{prop22} \\
B(y) & = & \frac{1}{4(D-1)H^2} \,
\Bigl[ -(4y-y^2) \gamma'(y) \, - \, (D-1)(2-y) \gamma(y) \Bigr] 
\;\; , \qquad \label{B2} \\
C(y) & = & \frac{1}{4(D-1)H^2} \,
\Bigl[ (2-y) \gamma'(y) \, - \, (D-1) \gamma(y) \Bigr] 
\;\; . \label{C2}
\end{eqnarray}

\subsection{Flat Spacetime Limit} 

One takes the flat space limit by letting $H$ approach zero with the 
physical time $t$ fixed, rather than the conformal time $\eta$:
\begin{eqnarray}
Flat \qquad & \Longrightarrow & \qquad 
H \longrightarrow 0 \quad \& \quad t \; {\rm fixed}
\;\; , \\
& \Longrightarrow & \qquad 
\eta \; = \; -\frac1{H} \, e^{-Ht} 
\; \longrightarrow \; 
-\frac1{H} + t
\;\; . \label{Heat}
\end{eqnarray}
We will express the flat space coordinate using the symbol $X^{\mu}$,
meaning $X^0 \equiv t$ and $X^i = x^i$. From the definition (\ref{y}) 
of $y(x;x')$, and relation (\ref{Heat}), we see that the flat space 
limiting forms for $y(x;x')$ and its derivatives are:
\begin{eqnarray}
Flat \qquad \Longrightarrow \qquad 
y(x;x') & \longrightarrow & H^2 \Delta X^2 
\;\; , \\
\frac{\partial y}{\partial x^{\mu}} \, 
\frac{\partial y}{\partial x^{\prime \nu}} 
& \longrightarrow & 
-4 H^4 \Delta X_{\mu} \, \Delta X_{\nu} 
\;\; , \label{dydy} \\
\frac{\partial^2 y}{\partial x^{\mu} \, \partial x^{\prime \nu}} 
& \longrightarrow & 
-2 H^2 \, \eta_{\mu\nu} \;\; . \label{ddy}
\end{eqnarray}
Here $\eta_{\mu\nu}$ is the Minkowski metric and the Minkowski 
coordinate interval is:
\begin{eqnarray}
\Delta X_{\mu} & \equiv &
\eta_{\mu\nu} (X - X')^{\nu}
\;\; , \\
\Delta X & \equiv & 
\sqrt{\Vert \vec{X} - \vec{X}'\Vert^2 - 
(\vert t - t'\vert - i\varepsilon)^2} 
\;\; .
\end{eqnarray} 

We also need asymptotic forms for functions of the parameter $\nu$ 
in the limit of small $H$ with the mass $m$ fixed. From expression 
(\ref{nu})
we see:
\begin{equation}
\nu \; \longrightarrow \; \frac{i m}{H} 
\;\; .
\end{equation}
Now use the Stirling approximation to obtain the asymptotic form 
of the ratios of Gamma functions which occur in the Laurent expansion 
(\ref{sol3}):
\begin{equation}
\frac{\Gamma(K + \frac12 \pm \nu)}{\Gamma(\frac12 \pm \nu)}
\; \longrightarrow \;
\frac{\Gamma(K + \frac12 \pm \frac{im}{H})}
{\Gamma(\frac12 \pm \frac{im}{H})}
\; \longrightarrow \; 
\Bigl(\pm \frac{im}{H}\Bigr)^K 
\;\; . \label{Gamas}
\end{equation}
By using (\ref{Gamas}) alternatively for $K = n + \frac{D}2$ and 
$K = n + 2$, we compute the flat space limit of (\ref{sol3}):
\begin{eqnarray}
\lefteqn{\lim_{flat} \gamma(y) \; = \;
\frac{D-1}{2} \, \Gamma \Bigl( \frac{D}2-1 \Bigr) \, 
\Gamma \Bigl( 2-\frac{D}2 \Bigr) \,
\frac{m^{D-2}}{(4 \pi)^{\frac{D}{2}}} } 
\nonumber \\
& & \hspace{3cm} 
\times \sum_{n=0}^{\infty} \Biggl[
\frac{(\frac14 m^2 \Delta X^2)^{n-\frac{D}2+1}}
{\Gamma(n + 2 -\frac{D}2) (n+1)!} \, - \,
\frac{(\frac14 m^2 \Delta X^2)^n}{\Gamma(n+1 + \frac{D}2) n!}
\Biggr] \;\; . \qquad \label{gammatilde}
\end{eqnarray}

We shall employ a tilde to denote the flat space limit of any 
quantity; for instance:
\begin{equation}
\widetilde{\gamma}(\Delta X^2) \; \equiv \; 
\lim_{flat} \gamma(y) 
\;\; .
\end{equation}
It is straightforward to check that expression (\ref{gammatilde}) 
obeys the equation:
\begin{eqnarray}
& \mbox{} & \hspace{-1cm} 
4 \, \Delta X^2 \, \widetilde{\gamma}''(\Delta X^2) \, + \,
2 (D+2) \, \widetilde{\gamma}'(\Delta X^2) \, - \,
m^2 \, \widetilde{\gamma}(\Delta X^2) 
\nonumber \\
& \mbox{} & \hspace{7.3cm} 
= \; \frac{D-1}2 \; 
\frac{\Gamma(\frac{D}2)}{\pi^{\frac{D}2}} \;
\frac1{\Delta X^{D}} 
\;\; . \qquad \label{inhomo}
\end{eqnarray}
The analogous quantity $\widetilde{\gamma}$ for Allen and
Jacobson is:
\begin{equation}
\widetilde{\gamma}_{\scriptscriptstyle \rm AJ}(\Delta X^2) 
\; = \; 
\widetilde{\gamma}(\Delta X^2) \, - \,
\frac{D-1}2 \, \frac{m^{D-2}}{(4 \pi)^{\frac{D}2}} \; 
\frac{\Gamma(\frac{D}2)}{(\frac14 m^2 \Delta X^2)^{\frac{D}2}} 
\;\; . \label{gammaAJ}
\end{equation}
This function obeys the homogeneous analogue of (\ref{inhomo}):
\begin{equation}
4 \, \Delta X^2 \; 
\widetilde{\gamma}''_{\scriptscriptstyle \rm AJ}(\Delta X^2) \, + \, 
2 (D+2) \; \widetilde{\gamma}'_{\scriptscriptstyle \rm AJ}(\Delta X^2) 
\, - \, m^2 \; 
\widetilde{\gamma}_{\scriptscriptstyle \rm AJ}(\Delta X^2) 
\; = \; 0 
\;\; . \label{homo}
\end{equation}

The flat space limit of the full propagator additionally requires 
expressions (\ref{prop22}-\ref{C2}) and the limiting forms 
(\ref{dydy}-\ref{ddy}):
\begin{eqnarray}
i \Bigl[ {}_{\mu} \widetilde{\Delta}_{\nu} \Bigr](x;x') & = &
\nonumber \\
& \mbox{} & \hspace{-1.8cm}
\eta_{\mu\nu} \; \widetilde{\gamma}(\Delta X^2) \, + \,
\frac2{D-1} \Bigl[ \, 
\eta_{\mu\nu} \, \Delta X^2 - \Delta X_{\mu} \; \Delta X_{\nu} 
\, \Bigr] \; \tilde{\gamma}'(\Delta X^2) 
\;\; . \label{flatprop}
\end{eqnarray}
Using relation (\ref{inhomo}) it is easy to check that 
(\ref{flatprop}) obeys the correct equation:
\begin{equation}
(\partial^2 - m^2) \;
i \Bigl[ {}_{\mu} \widetilde{\Delta}_{\nu} \Bigr](x;x') \; = \;
i \Bigl( \eta_{\mu\nu} - \frac{\partial_{\mu} \partial_{\nu}}
{\partial^2} \Bigr) \; \delta^D(x - x') 
\;\; . \label{nonlocal}
\end{equation}
The right hand side of (\ref{inhomo}) results in the non-local 
term on the right hand side of (\ref{nonlocal}) which is non-zero 
away from coincidence. The flat space limit of the Allen-Jacobson 
propagator is obtained by replacing $\widetilde{\gamma}$ by 
$\widetilde{\gamma}_{\scriptscriptstyle AJ}$ in (\ref{flatprop}). 
In contrast to the correct propagator, one can use relation 
(\ref{homo}) to show that the flat space limit of the Allen-Jacobson 
propagator obeys:
\begin{equation}
(\partial^2 - m^2) \;
i \Bigl[ {}_{\mu} \widetilde{\Delta}^{\scriptscriptstyle \rm AJ}_{\nu} 
\Bigr](x;x') \; = \; 
\frac{i}{m^2} \, \Bigl(\eta_{\mu\nu} \partial^2 - 
\partial_{\mu} \partial_{\nu} \Bigr) \; \delta^D(x -x') 
\;\; .
\end{equation}
The extra term in (\ref{gammaAJ}) has promoted the singularity from a
delta function to second derivatives of a delta function.

\subsection{Coincidence Limit} 

Obviously ``coincidence'' means $x^{\prime \mu} = x^{\mu}$. 
This implies:
\begin{itemize}
\item{All positive powers of $y$ vanish;}
\item{All $D$-dependent powers of $y$ vanish when using dimensional 
regularization; and}
\item{The tensors of the $y$-basis have the following limits:
\begin{equation}
\lim_{x' \rightarrow x} \frac{\partial y}{\partial x^{\mu}}
\frac{\partial y}{\partial x^{\prime \nu}} \; = \; 0 
\qquad , \qquad
\lim_{x' \rightarrow x} \frac{\partial^2 y}{\partial x^{\mu}
\partial x^{\prime \nu}} \; = \; 
-2 H^2 g_{\mu\nu} 
\;\; .
\end{equation}}
\end{itemize}

The first and second rules imply that only the $n=0$ terms from 
the two infinite sums of (\ref{sol3}) survive:
\begin{eqnarray}
\gamma(0) &\!\! = \!\!& 
\frac{D-1}{2} \, \frac{H^2}{m^2} \,
\frac{H^{D-2}}{(4\pi)^{\frac{D}{2}}} 
\label{coincsol} \\
& \mbox{} & 
\times \; \Bigg[ \,
\frac{\Gamma(D-1)}{\Gamma(\frac{D}{2}+1)} \, - \,
\Gamma(-\frac{D}{2}) \; \frac{\Gamma(\frac{D+1}{2}+\nu) \,
\Gamma(\frac{D+1}{2}-\nu)} {\Gamma(\frac12+\nu) \,
\Gamma(\frac12-\nu)} \, \Bigg]
\;\; . \nonumber
\end{eqnarray}
The first term inside the square brackets derives from our 
correction. The coincidence limit of the full propagator 
requires the third rule, in addition to relations (\ref{prop22}), 
(\ref{B2}) and (\ref{C2}):
\begin{equation}
\lim_{x' \rightarrow x} i \Bigl[ {}_{\mu} \Delta_{\nu} \Bigr](x;x') 
\; = \; 
\gamma(0) \, g_{\mu\nu} 
\;\; . \label{coinprop}
\end{equation}

\subsection{Massless Limit}

This is the most important limit because it gives the Lorentz gauge 
propagator. We therefore distinguish it with the convention that an
overline indicates the massless limit:
\begin{equation}
\lim_{m \rightarrow 0} \gamma(y) \; \equiv \; 
\overline{\gamma}(y) 
\qquad , \qquad
\lim_{m \rightarrow 0} i \Bigl[ {}_{\mu} \Delta_{\nu} \Bigr](x;x') 
\; \equiv \;
i \Bigl[ {}_{\mu} \overline{\Delta}_{\nu} \Bigr](x;x') 
\;\; .
\end{equation}
To take the massless limit we first expand the parameter $\nu$ 
(\ref{nu}) about its value for $m=0$:
\begin{equation}
\nu \; = \; \frac{D-3}{2} \, - \,
\frac{1}{D-3} \, \frac{m^2}{H^2} \, + \,
{\cal O} \Bigl( \frac{m^4}{H^4} \Bigr)
\; \equiv \; 
\frac{D-3}{2} \, + \, \Delta\nu
\;\; . \label{deltanu}
\end{equation}
We then substitute into form (\ref{sol3}) for $\gamma(y)$ and 
expand the $\nu$-dependent Gamma functions using the polygamma 
function \cite{GR}:
\begin{equation}
\psi(z) \; \equiv \;
\frac{d}{dz} \ln [ \, \Gamma(z) \, ] \; = \;
\psi(z-1) \, + \, \frac{1}{z-1}
\;\; . \label{psi}
\end{equation}
The result is:
\begin{eqnarray}
\overline{\gamma}(y) &\!\! = \!\!&
\frac{D-1}{2(D-3)} \,
\frac{H^{D-2}}{(4\pi)^{\frac{D}{2}}} \,
\Bigg\{
(D-3) \, \Gamma(\frac{D}{2}-1) \,
\left( \frac{4}{y} \right)^{\frac{D}{2}-1}
\label{masslessol} \\
& \mbox{} &
+ \, \sum_{n=0}^{\infty} \Bigg[ \, \frac{(n+1) \,
\Gamma(n+D-1)}{\Gamma(n+\frac{D}{2}+1)}
\nonumber \\
& \mbox{} & \hspace{-2.3cm}
\times \, \Big[ \, \psi(2-\frac{D}{2}) -
\psi(\frac{D}{2}-1) + \psi(n+D-1) - \psi(n+2) \, \Big]
\left( \frac{y}{4} \right)^n \Bigg]
\nonumber \\
& \mbox{} &
- \, \sum_{n=0}^{\infty} \Bigg[ \,
\frac{(n-\frac{D}{2}+3) \,
\Gamma(n+\frac{D}{2}+1)}{\Gamma(n+3)}
\nonumber \\
& \mbox{} & \hspace{-2.3cm}
\times \, \Big[ \, \psi(2-\frac{D}{2}) -
\psi(\frac{D}{2}-1) + \psi(n+\frac{D}{2}+1) -
\psi(n-\frac{D}{2}+4) \, \Big]
\left( \frac{y}{4} \right)^{n-\frac{D}{2}+2} \Bigg] 
\, \Bigg\}
\nonumber
\end{eqnarray}
Note that $\overline{\gamma}(y)$ would diverge without the extra
terms we have added to the corrected solution (\ref{sol}). The 
result for the massless limit of the propagator comes from simply 
replacing $\gamma(y)$ everywhere by $\overline{\gamma}(y)$ in 
expressions (\ref{prop22}), (\ref{B2}) and (\ref{C2}).

Additionally taking the flat space limit gives:
\begin{eqnarray}
\lim_{flat} \overline{\gamma}(y) & = &
\frac{D-1}{2} \,
\frac{\Gamma(\frac{D}{2}-1)}{4 \pi^{\frac{D}{2}}} \;
\frac{1}{\Delta X^{D-2}} 
\;\; , \label{flatsol} \\
\lim_{flat} i \Bigl[ {}_{\mu} \overline{\Delta}_{\nu} \Bigr](x;x') 
& = &
\frac{\Gamma(\frac{D}2)}{4 \pi^{\frac{D}2}} \, \Biggl[
\frac{\eta_{\mu\nu}}{D-2} \, + \, 
\frac{\Delta X_{\mu} \, \Delta X_{\nu}}{\Delta X^2} \Biggr] \, 
\frac1{\Delta X^{D-2}} 
\;\; .
\end{eqnarray}
Additionally taking the coincidence limit gives:
\begin{eqnarray}
\overline{\gamma}(0) &\!\!\! = \!\!\!&
\frac{D-1}{2(D-3)} \, 
\frac{H^{D-2}}{(4\pi)^{\frac{D}{2}}} \,
\frac{\Gamma(D-1)}{\Gamma(\frac{D}{2}+1)}
\nonumber \\
& \mbox{} &
\times \,\,
\left[ \, \psi(D-1) \, - \, \psi(2) \, - \, 
\psi(\frac{D}{2}-1) \, + \, \psi(2-\frac{D}{2}) \, \right]
\;\; . \label{masslesscoincsol} 
\end{eqnarray}
The coincidence limit of the massless limit of the propagator
comes from simply replacing $\gamma(0)$ by $\overline{\gamma}(0)$ 
in expression (\ref{coinprop}).

\subsection{$D=4$ Limit}

When using dimensional regularization in perturbative computations
involving scalar QED, we can take $D=4$ on any part of the result
which makes a finite contribution. This means one typically only 
needs to keep $D$ arbitrary in a few of the terms of expression 
(\ref{masslessol}) for $\overline{\gamma}(y)$. We have found the 
following expansions useful \cite{nctrpw1,PTW}:
\begin{eqnarray}
\lefteqn{\overline{\gamma}(y) \; = \;
\frac{D - 1}2 \, \Gamma\Bigl(\frac{D}2 \Bigr) \, 
\frac{H^{D-2}}{(4 \pi)^{\frac{D}2}} \, 
\Biggl\{ \frac{2}{D-2} \, 
\Bigl( \frac{4}{y} \Bigr)^{\frac{D}2-1} } 
\nonumber \\
& & \hspace{2.5cm}
+ \Bigl[ \, 1 + \frac2{D-2} - \frac2{D-4} \, \Bigr] 
\Bigl( \frac{4}{y} \Bigr)^{\frac{D}2-2} 
\nonumber \\
& & \hspace{1cm} 
+ \, \frac{\Gamma(\frac{D}2 + 1)}
{\Gamma(\frac{D}2) \, \Gamma(\frac{D}2 +2)} 
\Biggl[ \, \frac{\psi(2 - \frac{D}2) - \psi(\frac{D}2 -1) 
+ \psi(D-1) - \psi(2)}{(D-1) (D-3)} \, \Biggr] 
\nonumber \\
& & \hspace{2.5cm} 
+ \, \frac{\ln(\frac{y}4)}{(1 -\frac{y}4)^2} \, - \,
\ln\Bigl(\frac{y}4\Bigr) \, + \,
\frac{\frac{y}4}{1 - \frac{y}4} \, + \,
O\Bigl( \, (D-4) y \, \Bigr) \Biggr\} 
\;\; , \qquad
\end{eqnarray}
and:
\begin{eqnarray}
\overline{\gamma}(y) &\!\! = \!\!&
\frac{D - 1}2 \, \Gamma\Bigl( \frac{D}2 \Bigr) \,
\frac{H^{D-2}}{(4 \pi)^{\frac{D}2}} \, 
\Biggl\{ \frac2{D-2} \, 
\Bigl( \frac{4}{y} \Bigr)^{\frac{D}2-1} \, + \, 
\frac{\ln(\frac{y}{4})}{(1 - \frac{y}{4})^2} \, + \,
\frac{1}{1 - \frac{y}{4}} 
\nonumber \\ 
& \mbox{} & \hspace{7.8cm}
+ \, O(D-4) \Biggr\} 
\;\; . \qquad \label{massless4sol2}
\end{eqnarray}

\section{Response to a Point Source}

We begin by extracting a pure gauge term from our propagator 
(\ref{prop22}) using relations (\ref{B2}-\ref{C2}) and 
(\ref{gammaeqn2}):
\begin{eqnarray}
i \Bigl[ {}_{\mu} \Delta_{\nu} \Bigr](x;x') & = &
- \, \frac{1}{4(D-1)H^2} \,
\frac{\partial}{\partial x^{\mu}} \Bigg\{
\frac{\partial y}{\partial x^{\prime \hspace{0.05cm} \nu}}
\Big[ (4y-y^2) \, \gamma^{\prime}(y)
\nonumber \\
& \mbox{} & \hspace{5cm}
+ \, (D-1) (2-y) \, \gamma(y) \, \Big]
\Bigg\}
\nonumber \\
& \mbox{} &
+ \, \frac{1}{4(D-1)H^2} \,
\frac{\partial y}{\partial x^{\mu}} \,
\frac{\partial y}{\partial x^{\prime \hspace{0.05cm} \nu}} \,
\Big[ (4y-y^2) \, \gamma''(y) 
\nonumber \\
& \mbox{} & \hspace{1.5cm}
+ \, (D+2)(2-y) \, \gamma'(y) \, - \,
2(D-1) \, \gamma(y) \, \Big]
\;\; , \label{prop2} \\
& = & - \, \frac{1}{4(D-1)H^2} \,
\frac{\partial}{\partial x^{\mu}} \Bigg\{
\frac{\partial y}{\partial x^{\prime \hspace{0.05cm} \nu}}
\Big[ (4y-y^2) \, \gamma^{\prime}(y) 
\nonumber \\
& \mbox{} & \hspace{5cm}
 + \, (D-1) (2-y) \, \gamma(y) \, \Big]
\Bigg\}
\nonumber \\
& \mbox{} & \hspace{-0.7cm}
+ \, \frac{1}{4H^2} \, \Big[ \, 
\frac1{D-1} \, \frac{m^2}{H^2} \, \gamma(y) \, + \,
(2-y)A^{\prime}(y) \, - \, k \, \Big] \,
\frac{\partial y}{\partial x^{\mu}} \,
\frac{\partial y}{\partial x^{\prime \hspace{0.05cm} \nu}}
\; . \qquad \label{prop3}
\end{eqnarray}
Now recall that the retarded Green's function derives from the 
imaginary part of the propagator:
\begin{equation}
\Big[ {}_{\mu}G_{\nu}^{\rm ret}\Big](x;x^{\prime})
\; = \; -2 \theta(\eta - \eta') \;
{\rm Im} \Big\{ i \Big[ {}_{\mu}\Delta_{\nu}\Big](x;x') \Big\}
\;\; . \label{Gret}
\end{equation}
It only makes sense to extract the pure gauge term in the massless
limit, for which the theory is gauge invariant. When we  also take 
$D =4$, we conclude:
\begin{equation}
\lim_{D \rightarrow 4} \; (2 - y ) A'(y) \; = \;
\frac{H^2}{16\pi^2} \,
\left\{ - \frac{8}{y^2} \, + \, 2 \right\}
\;\; . \label{A'4}
\end{equation}
Hence the retarded Green's function is:
\begin{eqnarray}
\Big[ {}_{\mu}\overline{G}_{\nu}^{\rm ret} \Big](x;x^{\prime})
&\!\! = \!\!&
{\rm (pure \; gauge)}
\, - \,
\frac{2 \theta(\Delta\eta)}{4H^2} \,
\frac{H^2}{16\pi^2} \;
{\rm Im} \Bigg\{ \!\! - \frac{8}{(y + i\epsilon)^2} \Biggr\} \,
\frac{\partial y}{\partial x^{\mu}} \,
\frac{\partial y}{\partial x^{\prime \hspace{0.05cm} \nu}}
\nonumber \quad \\
&\!\! = \!\!&
{\rm (pure \; gauge)}
\, + \,
\frac{\theta(\Delta\eta)}{4\pi} \,
\delta^{\prime}(y) \,
\frac{\partial y}{\partial x^{\mu}} \,
\frac{\partial y}{\partial x^{\prime \hspace{0.05cm} \nu}}
\;\; , \label{Gret2}
\end{eqnarray}

The retarded Green's function determines the field response 
$A_{\mu}(x)$ due to the presence of a source $J^{\nu}(x)$:
\begin{equation}
A_{\mu}(x) \; = \; \int d^4 x' \;
\Big[ {}_{\mu}G_{\nu}^{\rm ret} \Big](x;x^{\prime}) \; 
J^{\nu}(x')
\;\; . \label{respeqn}
\end{equation}
We shall be concerned with a point source so that:
\begin{equation}
J^{\nu}(x) \; = \; q \int d\tau \;
{\dot \chi}^{\nu}(\tau) \; \delta^4 (x - \chi(\tau))
\;\; , \label{pointsource}
\end{equation}
where $\chi^{\mu}(\tau)$ is the geodesic of the source. 
Up to a pure gauge term, the point source response successively
equals:
\begin{eqnarray}
A_{\mu}(x) &\!\! = \!\!&
\frac{q}{4\pi} \int d^4 x' \;
\theta(\Delta\eta) \;
\delta^{\prime}(y) \;
\frac{\partial y}{\partial x^{\mu}} \,
\int d{\tau} \,
\frac{\partial y}{\partial x^{\prime \hspace{0.05cm} \nu}} \;
{\dot \chi}^{\nu}(\tau) \;
\delta^4 (x' - \chi(\tau))
\;\; , \nonumber \\
&\!\! = \!\!&
\frac{q}{4\pi} \int d{\tau} \;
\frac{d}{d\tau} \Big[ y(x ; \chi(\tau)) \Big] \,
\theta(\eta - \chi^0(\tau)) \;
\delta^{\prime}(y) \;
\frac{\partial y}{\partial x^{\mu}}
\;\; , \nonumber \\
&\!\! = \!\!&
\frac{q}{4\pi} \int_{-\infty}^{\frac{x^2}{\eta^2}} \;
\delta^{\prime}(y) \;
\frac{\partial y}{\partial x^{\mu}}
\;\; , \nonumber \\
&\!\! = \!\!&
\frac{q}{4\pi} \,
\left[ - \frac{\partial}{\partial y} \,
\frac{\partial y}{\partial x^{\mu}} \right]_{y=0}
\;\; . \label{response}
\end{eqnarray}
For a particle at rest on the conformal origin the length 
function is:
\begin{equation}
\chi^{\mu}(\tau) \, = \, \tau \, \delta^{\mu}_{~0}
\quad \Longrightarrow \quad
y(x ; \chi(\tau)) \, = \,
\frac{1}{\eta\tau} \,
\Big[ x^2 \, - \, (\eta - \tau)^2 \Big]
\;\; , \label{geodesic}
\end{equation}
where $\tau$ represents the conformal time of the source 
and $\eta$ the conformal time of the observation event. 
The relevant derivatives for our computation are:
\begin{equation}
\frac{\partial y}{\partial \eta} \, = \,
- \frac{x^2}{\eta^2 \tau} - \frac{1}{\tau} +
\frac{\tau}{\eta^2}
\quad , \quad
\frac{\partial y}{\partial x^i} \, = \,
\frac{2x^i}{\eta\tau}
\quad ; \quad
\frac{\partial y}{\partial\tau} \Bigg\vert_{y=0} \, = \,
\frac{2x}{\eta (\eta-x)}
\;\; , \label{derivatives}
\end{equation}
leading to the final expressions for the response:
\begin{eqnarray}
A_0(x) &\! = \!&
- \frac{q}{4\pi} \, \left[
\frac{x^2 + \eta^2 - \eta x}{\eta x \, (\eta - x)} \right]
\;\; , \label{response0} \\
A_i(x) &\! = \!&
\frac{q}{4\pi} \;
\frac{x^i}{x \, (\eta - x)}
\;\; , \label{responsei}
\end{eqnarray}
and invariant field strength tensor:
\begin{equation}
F_{0i} \, = \,
- \frac{q}{4\pi} \, \frac{x^i}{x^3}
\qquad , \qquad
F_{ij} \, = \, 0
\;\; . \label{Fmunu}
\end{equation}
This is the correct result in our coordinate system \cite{RPW} so
we conclude that our propagator provides the proper response.

\section{Epilogue}

We have worked out the Feynman propagator for a massive vector 
field on $D$-dimensional de Sitter space. This involved repairing
a minor error in the classic analysis of Allen and Jacobson 
\cite{allenjacobson}. They failed to include a term which is 
non-zero even away from coincidence and is required to make the 
right hand side of the propagator equation (\ref{propeqn}) respect 
the transversality of the left hand side. Without this term, the
propagator diverges in the massless limit and also fails to agree
with the known result in the flat space limit. The correction is
nonetheless small, and much of the solution is simply the transcription
to our notation of the work of Allen and Jacobson.

Notation deserves comment as well. It is traditional to represent de
Sitter invariant bi-vectors such as the propagator with scalar functions
of the geodesic length, $\ell(x;x')$, times a tensor basis formed from
gradients of $\ell(x;x')$ on $x^{\mu}$ and $x^{\prime\mu}$, and the
parallel transport matrix $[\mbox{}_{\mu} g_{\nu}](x;x')$. We have 
instead chosen to use a function $y(x;x')$ of the geodesic length which 
has a simple expression (\ref{y}) in the conformal coordinate system 
most often used for explicit computations. We have also taken our tensor 
basis to consist of the de Sitter invariant first and mixed second 
derivatives of $y(x;x')$. This greatly facilitates computations in 
theories with derivative interactions such as scalar quantum 
electrodynamics \cite{nctrpw1,PTW} and gravity. A few simple identities 
suffice for all contractions (\ref{ID1}-\ref{ID3}) and covariant 
derivatives (\ref{ID4}-\ref{ID5}) of these tensors.

Finally, we demonstrated that the retarded Green's function inferred 
from our propagator (in the massless limit) correctly reproduces the
known classical response to a point charge on the open sub-manifold.
This is an issue because the peculiar causal structure and linearization 
instability of de Sitter space invalidate some of the familiar features 
of flat space electrodynamics \cite{RP,nctrpw2,BK1,BK2}. There also seems 
to be a poorly understood obstacle to imposing certain gauges in gravity 
and electrodynamics \cite{AM,RPW}. In particular, it has been shown that 
de Sitter-Feynman gauge leads to on-shell singularities in the one loop 
scalar self-mass-squared of scalar quantum electrodynamics \cite{KW}. It
is reassuring that our propagator leads to no such problem \cite{PTW,KW2}.

\vspace{1cm}

\centerline{\bf Acknowledgements}

This work was partially supported by the European Social fund
and National resources Y$\Pi$E$\Pi\Theta$-PythagorasII-2103,
by European Union grants FP-6-012679 and MRTN-CT-2004-512194,
by NSF grant PHY-0244714, and by the Institute for Fundamental
Theory at the University of Florida.


\vspace{1cm}

\begin{thebibliography}{99}

\bibitem{AP} A. Proca, C. R. Acad. Sci. Paris {\bf 202} (1936) 1490.

\bibitem{CW} S. R. Coleman and E. Weinberg, Phys. Rev. {\bf D7} (1973) 1888.

\bibitem{IMO} T. Inagaki, T. Muta and S. D. Odintsov, Prog. Theor. Phys.
Suppl. {\bf 127} (1997) 93, \\
{\bf arXiv:}hep-th/9711084.

\bibitem{nctrpw1} T. Prokopec, N. C. Tsamis and R. P. Woodard,
                  Annals Phys. {\bf 323} (2008) 1324, \\
                  {\bf arXiv:}0707.0847.
                  

\bibitem{PTW} T. Prokopec, N. C. Tsamis and R. P. Woodard, 
              Class. Quant. Grav. {\bf 24} (2007) 201, \\
              {\bf arXiv:}gr-qc/0607094.

\bibitem{allenjacobson} B. Allen and T. Jacobson,
                        Commun. Math. Phys. {\bf 103} (1986) 669.

\bibitem{peters} P. C. Peters, J. Math. Phys. {\bf 10} (1969) 1216.

\bibitem{KW} E. O. Kahya and R. P. Woodard, Phys. Rev. {\bf D72} (2005)
104001, \\ {\bf arXiv:}gr-qc/0508015.

\bibitem{allenfolacci} B. Allen and A. Folacci,
                       Phys. Rev. {\bf D35} (1987) 3771.

\bibitem{OW1} V. K. Onemli and R. P. Woodard, Class. Quant. Grav. \textbf{19}
(2002) 4607, \\
{\bf arXiv:}gr-qc/0204065.

\bibitem{OW2} V. K. Onemli and R. P. Woodard, Phys. Rev. D \textbf{70} (2004)
107301, \\
{\bf arXiv:}gr-qc/0406098.

\bibitem{GR} I. S. Gradshteyn and I. M. Ryzhik, {\it Table of Integrals, 
Series and Products}, 4th ed. (Academic Press, New York, 1965).

\bibitem{RPW} R. P. Woodard, ``de Sitter Breaking in Field Theory,'' in
{\it DESERFEST: A Celebration of the Life and Works of Stanley Deser}, ed.
J. T. Liu, M. J. Duff, K. S. Stelle and R. P. Woodard (World Scientific, 
Sinapore, 2006) pp. 339-351, \\
{\bf arXiv:}gr-qc/0408002.

\bibitem{RP} R. Penrose, in {\it Relativity, Groups and Topology, Les Houches
1963}, ed. C. DeWitt and B. DeWitt (Gordon and Breach, New York, 1964).

\bibitem{nctrpw2} N. C. Tsamis and R. P. Woodard,
                  Commun. Math. Phys. {\bf 162} (1994) 217.

\bibitem{BK1} J. Bicak and P. Krtous, Phys. Rev. {\bf D64} (2001) 124020, \\
{\bf arXiv:}gr-qc/0107078.

\bibitem{BK2} J. Bicak and P. Krtous, J. Math. Phys. {\bf 46} (2005) 102504, \\
{\bf arXiv:}gr-qc/0602009.

\bibitem{AM} I. Antoniadis and E. Mottola, J. Math. Phys. {\bf 32} (1991) 1037.

\bibitem{KW2} E. O. Kahya and R. P. Woodard, Phys. Rev. {\bf D74} (2006)
              084012, \\
              {\bf arXiv:}gr-qc/0608049.

\end{thebibliography}
\end{document}